# A NEW ARCHITECTURE OF A UBIQUITOUS HEALTH MONITORING SYSTEM:

## A Prototype Of Cloud Mobile Health Monitoring System

Abderrahim BOUROUIS[1], Mohamed FEHAM[2] and Abdelhamid BOUCHACHIA[3]

[1]STIC laboratory, Abou-bekr BELKAID University,Tlemcen,Algeria
a_bourouis@mail.univ-tlemcen.dz

[2]STIC laboratory, Abou-bekr BELKAID University,Tlemcen,Algeria
m_feham@mail.univ-tlemcen.dz

[3]Research Group,Software Engineering and Soft Computing,University of Klagenfurt, Austria
hamid@isys.uni-klu.ac.at

**Abstract**
Wireless Body Area Sensor Networks (WBASN) is an emerging technology which uses wireless sensors to implement real-time wearable health monitoring of patients to enhance independent living. In this paper we propose a prototype of cloud mobile health monitoring system. The system uses WBASN and Smartphone application that uses cloud computing, location data and a neural network to determine the state of patients.

## 1.Introduction & Related Works

Recent technological advances in sensors facilitate wireless sensor networks that are deeply embedded in their native environments. Wireless sensor networks are highly suitable for many applications, such as habitat monitoring [1], machine health monitoring and guidance, traffic pattern monitoring and navigation, plant monitoring in agriculture [2], and infrastructure monitoring.

The current technological and economic trends enable new generations of wireless sensor networks with more compact and lighter sensor nodes, more processing power, and more storage capacity. In addition, the ongoing proliferation of wireless sensor networks across many application domains will result in a significant cost reduction.

One of the most promising application domains is health monitoring [3], and within health care, WBANs in particular are emerging as promising enabling technologies to implement m-health. A WBASN for health monitoring consists of multiple sensor nodes that can measure and report the user's physiological state. These sensors are strategically placed on the human body. The exact location of the sensor nodes on the human body depend on the sensor type, size and weight. Sensors can be worn as stand-alone devices or can be built into jewelry, applied as tiny patches on the skin, hidden in the user's clothes or shoes, or even implanted in the user's body. Each node in the WBASN is typically capable of sensing, sampling, processing, and wirelessly communicating one or more physiological signals.

The exact number and type of physiological signals to be measured, processed, and reported, depends on end-user application and may include many physiological sensors (ECG, EMG, EEG, SPO2…).

In addition to these sensors, a WBASN for health monitoring may include sensors that can help to determine the user's location, discriminate among the user's states (e.g., lying, sitting, walking, running), or estimate the type and level of the user's physical activity.

In this paper, we describe new Mobile Health Monitoring system architecture, it uses a wireless body area networks (WBASN) to collect and send data to the cloud server through GPRS/UMTS [4].

The system uses a cloud service to extract patient data information. These sensory parameters are fed into a neural network engine running as a cloud service that fuses information from multiple disparate sensors to determine whether the patient is in a "danger state".

The merger of mobile computing and cloud resources is one of the popular areas of research today. There are several application including mobile social application [5], healthcare application [6, 7], gamming [8], and localization [9] that leverage cloud resources and machine learning algorithms to design efficient systems. The tradeoff lies in between energy consumption and computability, and in some cases best of both words can be leveraged.

Choi et al. [10] proposed a system for ubiquitous health monitoring in the Bedroom via a Bluetooth Network and Wireless LAN. The information gathered from sensors connected to the patient's bed is transmitted to a monitoring station outside of the room where the data is processed and analyzed.

Using the technologies of wireless body area networks (WBAN), Jovanov et al. [11] presented a Wearable health systems using WBAN for patient monitoring. The first level consists of physiological sensors, the second level is the personal server, and the third level is the health care servers and related services.

Another example from Dai et al. [12], is a wireless physiological multi-parameter monitoring system based on mobile communication networks; this system monitors vital signs such as ECG, SP02, body temperature and respiration. Data is transmitted via mobile communications networks to a mobile monitoring station and then to the hospitals central management system where, again, the data must be reviewed and interpreted by a physician or other medical personnel.

The rest of the paper is organized as follows: Section 2 provides the system architecture and the functions of major components; in Section 3 we describe our research contribution and implementation. Section 4 summarizes and concludes this paper.

## 2.System Architecture

The wireless body area sensor network for ubiquitous health monitoring contains three components (Tier 1, 2 & 3):
Typically, a WBASN will form the lowest tier (Tier 1) of a multitiered medical information system for health monitoring. Figure 1 illustrates general system architecture of a medical monitoring information system that includes a personal server at Tier 2 and a series of medical servers in the cloud at Tier 3. The exact system architecture and the number of system tiers depend predominantly on target applications, available infrastructure, and type and number of users.
The WBASN can include heart sensor, motion sensors…
For example similar system can be used for monitoring of cardiac patients during a rehabilitation period at home. The heart sensor can operate in multiple modes reporting either a raw ECG signal (from one or multiple channels), time-stamped heart beats, or averaged heart rate over a certain period of time. The motion sensors, each equipped with a 3D accelerometer, can also operate in several modes, reporting either (1) raw acceleration signals for x-, y-, and z-axes, (2) extracted features (e.g., times tamped steps or phases of a step), or (3) an estimated level of activity. The sensor nodes (together with a network coordinator), attached to a personal server, compose the WBASN. Upon configuration, the WBASN continually performs sensing, sampling and signal processing. Sensors wait for command and control messages from the WBASN coordinator and report continual sensor readings or events of interest as they occur.

Tier 2 encompasses the personal server, which is responsible for a number of tasks, providing a transparent interface to the wireless sensor nodes, an interface to the user, and an interface to the medical server. The interface to the WBASN includes network configuration and management. Network configuration encompasses the following tasks: sensor node registration (type and number of sensors), initialization (e.g., specifying sampling frequency and mode of operation), customization (e.g., running user-specific calibration or user-specific signal processing procedure upload) and setup of a secure communication security key exchange. Once the WBASN network is configured, the personal server manages the network and takes care of channel sharing, time synchronization, data retrieval and processing, and fusion of the data. Based on synergy of information from multiple physiological, location, activity, and environmental sensors, the personal server can determine users' states and their health status; in addition, the personal server can provide feedback through a user-friendly and intuitive graphical or audio user interface.

The personal server is often already integrated with sensors; such as accelerometer to determine mobility and global positioning system (GPS) for location, which makes them attractive for a fully integrated wearable mobility monitoring system.

Finally, if a communication channel to the cloud is available, the personal server can establish a secure link to the cloud part and send condensed or detailed reports about users' health status. These reports can be processed, displayed, and integrated into users' medical records. However, if a link between the personal server and the cloud is not available, the personal server should be able to store the data locally and initiate uploads when a link becomes available. Depending on the use scenario, the personal server runs on a smart phone (as illustrated in Figure 1.

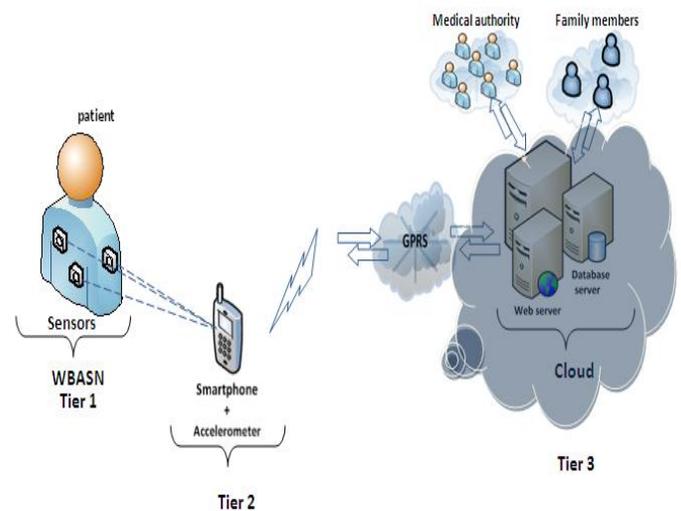

Fig.1 The system architecture

Tier 3 includes a cloud medical server accessed via the Internet.
In addition to the medical server in the cloud, the last tier may encompass other servers, such as informal caregivers, commercial health care providers, and even emergency services. The medical server keeps electronic medical records of registered users and provides various services to the users, medical personnel and informal caregivers. It is the responsibility of the medical server to authenticate users, accept health monitoring session uploads, format and insert the session data into corresponding medical records, analyze the data patterns, recognize serious health anomalies in order to contact emergency caregivers, and forward new instructions to the users, such as physician prescribed exercises. Patients' physicians can access the data from their offices via the Internet and examine the data to ensure that the patients are within expected health metrics (in terms of heart rate, blood pressure, activity) and that they are responding to a given treatment or performing prescribed exercises. A server agent may inspect the uploaded data and create an alert in the case of a potential medical condition.

The large amount of data collected through these services can also be utilized for knowledge discovery through data mining. Integration of the collected data into research databases and quantitative analysis of conditions and patterns likely will prove invaluable to researchers trying to link symptoms and diagnoses with historical changes in health status, physiological data, or other parameters (e.g., gender, age, weight).

## 3.Research contribution & Implementation

The research contributions in our architecture are:

1. We use a mobile cloud service to extract patient data information. These sensory parameters are fed into a neural network engine running as a cloud service that fuses information from WBASN to determine whether the patient is in a "danger state".

2. We propose a novel approach for smart and real-time analysis of sensors data using Smartphone and cloud back propagation Neural Network.

3. Our architecture uses rapid prototyping on the mobile platform (Smartphone) with Android operating system as the personal server (tier 2). It is developed in Python using Scripting Layer for Android (SL4A), the main benefit of this type of programming philosophy is that the development time of applications is radically decreased and it is a good support for client-server communications.

4. We propose a hybrid location system to determine the location of patient: It uses Smartphone integrated GPS if the patient is in GPS coverage zone, if not it collects Location Area Identifications such as: Mobile Country Code (MCC), Mobile Network Code (MNC), Location Area Code (LAC) and cell identification (CI) from cellular network to calculate the current location.

We have implemented a fully functional prototype; the current sensor being used in tier1 of our architecture is the Nonin 4100 Bluetooth pulse oximeter. It measures blood-oxygen saturation levels (Sp02) as well as heart rate (HR) [13].

We have considered a Sumsung Galaxy S as hardware platforms of tier 2 based on Android 2.3.3 operating System. We have developed a Python application for Android using Scripting Layer for Android (SL4A) based on HTML GUI, which gathers the sensor data from Nonin sensor (SpO2, HR, body temperature) (Figure 2). Smartphone uses Python APIs (Application Program Interfaces) to manage Bluetooth connections. Once data is received, it uses algorithms to compare currents data with the previous; if it detects changes it forwards that data to the cloud.

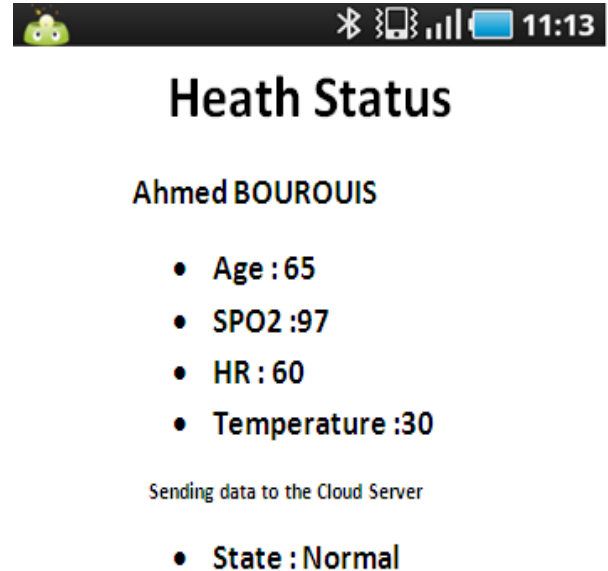

Fig.2 Use of a simple HTML and Python to display health status

Our system uses a neural network technique in the cloud to mine data and predict health risk from knowledge of the bio-signal sensor data.

Data from the Smartphone is fed into a neural network engine which uses a back propagation algorithm to fuse the data from the sensors into a single binary result (whether the state is a normal or not). We use a single layer (with one hidden layer) perception based neural network (Figure 3). The network is initially trained using samples collected from our application. We use accepted norms to determine whether a piece of data can be classified as an anomaly.

The neural network engine fuses data from four inputs and uses one layer of hidden perceptions to calculate the sigmoid function.

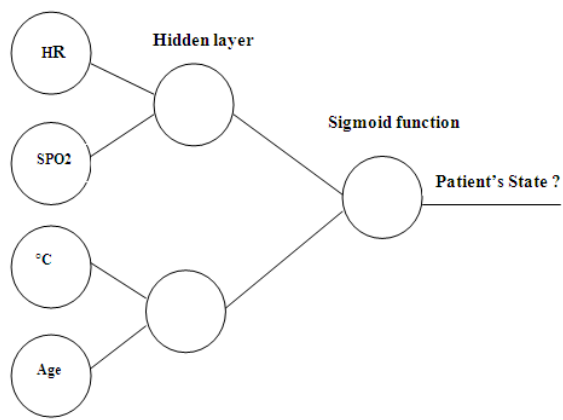

Fig.3 The overview of the neural network engine

The neural network engine was implemented using the Encog Neural Network Framwork for .Net [14]. There are two weighted synapse: connecting input layer and hidden layer and connecting hidden layer to output layer. For training the neural network we used data corresponding to 540 instances. We assume that if value of the sigmoid output is greater than 0.5, the patient is in an anomaly state otherwise he is not.

Across trials, the median accuracy is around 87% when only three inputs are used. However, the accuracy improves to 94% when the number of inputs is increased to 4. Note that this comes at the cost of longer time for convergence (Figure 4).

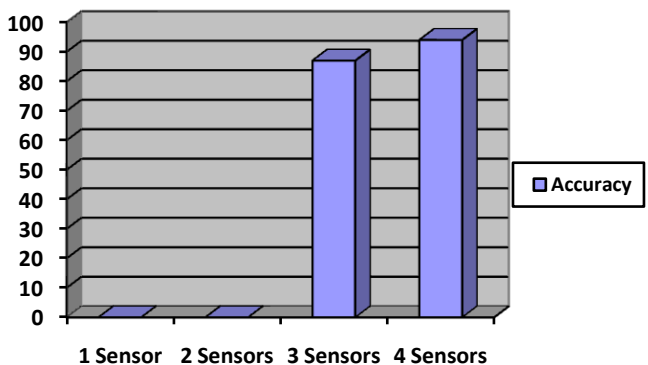

Fig.4 The accuracy of cloud neural network using 3 and 4 inputs

After processing the information, family members or medical authority can identify the real-time health status of the patient through a web application (figure 5). Once an abnormal situation is detected, an alert signal is sent, giving the medical professional an idea of the patient's health status and its current location in the case of an emergency.

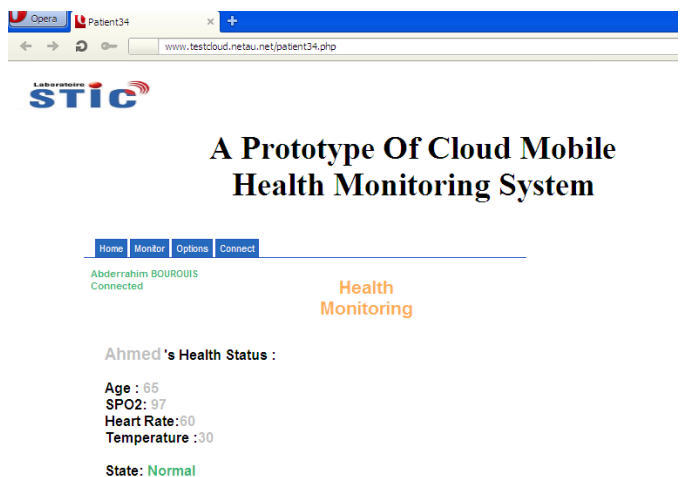

Fig.5 Web interface of the prototype

## 4.Conclusion

In conclusion, a new architecture of cloud mobile health monitoring system is presented for continuous monitoring of patients. The system provides the architecture for collecting, gathering and analyzing data from a number of biosensors using WBASN, Personal server and cloud medical server. The designed prototype system monitors location and health status through the use of a Nonin 4100 Bluetooth pulse oximeter as WBASN and Sumsung Galaxy S based on Android Python application. By providing capabilities for processing of the measurements and user I/O in the cloud part, the family members or medical authority are alerted in a timely manner when the state of patient changes for the worse.

## Acknowledgements

We would like to thank the STIC Laboratory personal Pr. Mohamed FEHAM for their publication support.